\documentclass[twocolumn,aps,nofootinbib,superscriptaddress,10pt]{revtex4}

\usepackage{units}
\usepackage{dsfont}

\usepackage{amsmath}
\usepackage{amssymb}
\usepackage{amsthm}

\newcommand{\comment}[1]{}

\newcommand{\abs}[1]{\ensuremath{|#1|}}
\newcommand{\absbig}[1]{\ensuremath{\big|#1\big|}}

\newcommand{\norm}[2]{\ensuremath{|\!|#1|\!|_{#2}}} 
\newcommand{\normbig}[2]{\ensuremath{\big|\!\big|#1\big|\!\big|_{#2}}}

\newcommand{\tr}{\textnormal{tr}}
\newcommand{\trace}[1]{\ensuremath{\tr (#1)}}
\newcommand{\Trace}[1]{\ensuremath{\tr \left( #1 \right)}}
\newcommand{\tracebig}[1]{\ensuremath{\tr \big( #1 \big)}}
\newcommand{\ptr}[1]{\textnormal{tr}_{\textnormal{\tiny #1}}}
\newcommand{\ptrace}[2]{\ensuremath{\ptr{#1} (#2)}}

\newcommand{\supp}[1]{\textnormal{supp}\, \{ #1 \}}

\newcommand{\idx}[2]{{#1}_{\textnormal{\tiny #2}}}

\newcommand{\ket}[1]{| #1 \rangle}

\newcommand{\braket}[2]{\langle #1 | #2 \rangle}

\newcommand{\bracket}[3]{\langle #1 | #2 | #3 \rangle}

\newcommand{\proj}[2]{| #1 \rangle\!\langle #2 |}

\newcommand{\iso}{\cong}

\newcommand{\kron}{\otimes}
\newcommand{\eps}{\varepsilon}

\newcommand{\h}{\ensuremath{\mathcal{H}}}
\newcommand{\hi}[1]{\ensuremath{\mathcal{H}_{\textnormal{\tiny #1}}}}
\newcommand{\hA}{\hi{A}}
\newcommand{\hB}{\hi{B}}
\newcommand{\hC}{\hi{C}}

\newcommand{\hAB}{\hi{AB}}
\newcommand{\hABC}{\hi{ABC}}

\newcommand{\id}{\ensuremath{\mathds{1}}}
\newcommand{\idi}[1]{\ensuremath{\mathds{1}_{\textnormal{\tiny #1}}}}
\newcommand{\idA}{\idi{A}}
\newcommand{\idB}{\idi{B}}

\newcommand{\positive}[1]{\ensuremath{\mathcal{P}(#1)}}
\newcommand{\states}[1]{\ensuremath{\mathcal{S}(#1)}}

\newcommand{\subnormstates}[1]{\ensuremath{\mathcal{S}_{\leq}(#1)}}

\newcommand{\statesB}{\states{\hB}}

\newcommand{\rhot}{\ensuremath{\tilde{\rho}}}

\newcommand{\rhoB}{\ensuremath{\idx{\rho}{B}}}
\newcommand{\rhoC}{\ensuremath{\idx{\rho}{C}}}

\newcommand{\rhoAB}{\ensuremath{\idx{\rho}{AB}}}
\newcommand{\rhotAB}{\ensuremath{\idx{\tilde{\rho}}{AB}}}
\newcommand{\rhoAC}{\ensuremath{\idx{\rho}{AC}}}

\newcommand{\rhoABC}{\ensuremath{\idx{\rho}{ABC}}}

\newcommand{\sigmaB}{\ensuremath{\idx{\sigma}{B}}}

\newcommand{\hh}[4]{\ensuremath{H_{#1}^{#2}(\textnormal{#3})_{#4}}}
\newcommand{\chh}[5]{\ensuremath{H_{#1}^{#2}(\textnormal{#3}|\textnormal{#4})_{#5}}}

\newcommand{\hmin}[2]{\hh{\textnormal{min}}{}{#1}{#2}} 
\newcommand{\hmineps}[3]{\hh{\textnormal{min}}{#1}{#2}{#3}}
\newcommand{\chmin}[3]{\chh{\textnormal{min}}{}{#1}{#2}{#3}}
\newcommand{\chmineps}[4]{\chh{\textnormal{min}}{#1}{#2}{#3}{#4}}
\newcommand{\hmaxeps}[3]{\hh{\textnormal{max}}{#1}{#2}{#3}}
\newcommand{\chmax}[3]{\chh{\textnormal{max}}{}{#1}{#2}{#3}}
\newcommand{\chmaxeps}[4]{\chh{\textnormal{max}}{#1}{#2}{#3}{#4}}

\newcommand{\hvn}[2]{\hh{}{}{#1}{#2}} 
\newcommand{\chvn}[3]{\chh{}{}{#1}{#2}{#3}}
\newcommand{\halpha}[2]{\hh{\alpha}{}{#1}{#2}} 
\newcommand{\chalpha}[3]{\chh{\alpha}{}{#1}{#2}{#3}}
\newcommand{\hx}[3]{\hh{#1}{}{#2}{#3}} 
\newcommand{\chx}[4]{\chh{#1}{}{#2}{#3}{#4}}

\newcommand{\epsball}[2]{\ensuremath{\mathcal{B}^{#1}(#2)}}

\theoremstyle{plain}
\newtheorem{lemma}{Lemma}
\newtheorem{theorem}[lemma]{Theorem}

\theoremstyle{definition}
\newtheorem{definition}{Definition}

\newtheorem{remark}[lemma]{Remark}

\begin{document}

\title{A Fully Quantum Asymptotic Equipartition Property}

\date{May 12, 2009}

\author{Marco \surname{Tomamichel}}
\email[]{marcoto@phys.ethz.ch}
\affiliation{Institute for Theoretical Physics, ETH Zurich, 8093
  Zurich, Switzerland.}
\author{Roger \surname{Colbeck}}
\email[]{colbeck@phys.ethz.ch}
\affiliation{Institute for Theoretical Physics, ETH Zurich, 8093
  Zurich, Switzerland.}
\affiliation{Institute of Theoretical Computer Science, ETH Zurich, 8092
  Zurich, Switzerland.}
\author{Renato \surname{Renner}}
\email[]{renner@phys.ethz.ch}
\affiliation{Institute for Theoretical Physics, ETH Zurich, 8093
  Zurich, Switzerland.}

\begin{abstract}
The classical asymptotic equipartition property is the statement
that, in the limit of a large number of identical repetitions of a
random experiment, the output sequence is virtually certain to come
from the typical set, each member of which is almost equally likely.
In this paper, we prove a fully quantum generalization of this
property, where both the output of the experiment and side
information are quantum. We give an explicit bound on the convergence,
which is independent of the dimensionality of the side information.
This naturally leads to a family of
R\'enyi-like quantum conditional entropies, for which the von
Neumann entropy emerges as a special case.
\end{abstract}

\maketitle

\section{Introduction}

In this paper, we prove a fully quantum version of
the asymptotic equipartition property (AEP). While the classical AEP
applies to a random experiment with classical outcomes, here we
consider the generalization to experiments that require a quantum
description.  Our version of the AEP then refers to typical
properties of outcomes of the experiment relative to some side
information, i.e., additional information obtained in the process of
the experiment.  We call it \emph{fully quantum} because the outcomes
as well as the side information may be quantum systems.  We note here
that in classical versions of the AEP the side information is not
usually described explicitly, but is already included in the
specification of the distribution of the experimental outcomes (i.e.,
one considers the probability distribution conditioned on the side
information). This is not possible in a fully quantum context, where
the side information may be entangled with the outcome and a quantum
analogue of conditional probability distributions cannot be defined.

We will first discuss the classical AEP and rewrite it in a form that
can then be generalized to the fully quantum setting. We sketch a possible proof
of the classical version and our proof of the fully quantum AEP will follow similar lines.

\subsection{Classical AEP}
\label{sec:classical}

The AEP (cf.\ Theorem 3.1.1 in \cite{cover91}) is central to classical
information theory because it establishes the Shannon
entropy\footnote{We use $\log$ to denote the binary logarithm.},
\begin{equation*}
\hvn{$X$}{} = -\sum_{x \in \mathcal{X}} P(x) \log P(x) ,
\end{equation*} 
as the relevant quantity for various problems involving independent
and identically distributed (i.i.d.) random variables.  It is a direct
consequence of the weak law of large numbers and states that, for
large enough $n$, the outcome of a random experiment given by an
i.i.d.\ sequence $X^n = (X_1, X_2, \ldots, X_n) \in
\mathcal{X}^{\times n}$ of random variables distributed according to a
probability distribution $P$ on a set $\mathcal{X}$ will almost
certainly be in a set of approximately $2^{n \hvn{$X$}{}}$ typical
events that each occur with a probability close to $2^{-n
  \hvn{$X$}{}}$.

Consider, for example, the problem of source compression. There, one
asks for the number of bits needed to store the outcome of the above
random experiment.  In typical information theoretic applications, we
tolerate a small probability of failure. The AEP tells us that if we
ignore non-typical events, we are almost certain not to have an error.  
We thus only need $n \hvn{$X$}{}$ bits to store the whole sequence, i.e.\ $\hvn{$X$}{}$
bits per element.

This can alternatively be formulated in terms of the entropies
\begin{eqnarray*}
\hx{\infty}{$X$}{} &:=& - \log \max_{x \in \mathcal{X}} P(x)\quad \textnormal{and}\\
\hx{0}{$X$}{} &:=& \log \absbig{ \{ x \in \mathcal{X} : P(x) > 0 \} } .
\end{eqnarray*}
We also informally introduce smooth min- and max-entropies denoted
$\hmineps{\eps}{$X$}{}$ and $\hmaxeps{\eps}{$X$}{}$, which will be
defined precisely in Section \ref{sec:def}. The smooth min-entropy is
constructed by ignoring the most probable events in $\mathcal{X}$ up
to total probability $\eps$ and taking $H_{\infty}$ of the remaining
distribution.  Similarly, the smooth max-entropy ignores the least
probable events and is closely related to $H_0$.  In terms of these
entropies, the AEP is equivalent to the relations
\begin{eqnarray}
&&\lim_{\eps \to 0} \lim_{n \to \infty} \frac{1}{n} \hmineps{\eps}{$X^n$}{} = \hvn{$X$}{} \quad \textnormal{and}
\label{eqn:aep} \\
&&\lim_{\eps \to 0} \lim_{n \to \infty} \frac{1}{n} \hmaxeps{\eps}{$X^n$}{} = \hvn{$X$}{}. \label{eqn:aep2} 
\end{eqnarray}
These relations have been generalized to the case of conditional
entropies (for a non-asymptotic version, see \cite{holenstein06}).

Returning to the example of source compression makes clear the second
of these relations.  In order to store (with certainty) the outcome of
a single random experiment, one needs
$\hx{0}{$X$}{}$ bits.  Furthermore, if one tolerates a small
probability of failure, only roughly $\hmaxeps{\eps}{$X$}{}$ bits are
required.  On the other hand, in the case of a large sequence of i.i.d.\ random
variables, the AEP tells us that $\hvn{$X$}{}$ bits are needed for
each element of the sequence, and hence relation \eqref{eqn:aep2}
follows.  A similar argument can be made to illustrate relation \eqref{eqn:aep} 
using randomness extraction \cite{renner04}.

We now sketch a proof of the AEP \eqref{eqn:aep}.
We use the R\'enyi entropies \cite{renyi61}
\begin{equation}
\halpha{$X$}{} := \frac{1}{1 - \alpha} \log \sum_{x \in \mathcal{X}} P(x)^\alpha, 
\quad \alpha \in (0, 1) \cup (1, \infty) ,
\label{eqn:renyi}
\end{equation}
for which $H_\infty$ ($\alpha \!\to\! \infty$), $H_0$ ($\alpha \!\to\! 0$) and the Shannon entropy ($\alpha \!\to\! 1$) are defined as limits.  Furthermore, the entropies $H_\alpha$ are
monotonically decreasing in $\alpha$ and, as shown in \cite{renner04},
the R\'enyi entropies with $\alpha > 1$ are close to the smooth
min-entropy in the sense that
\begin{equation}
\hmineps{\eps}{$X$}{} \geq \halpha{$X$}{} - \frac{1}{\alpha - 1} \log \frac{1}{\eps}, \qquad \alpha > 1
\label{eqn:class-entropy-ineq}
\end{equation}
while those with $\alpha < 1$ are close to the smooth max-entropy.
Note that the error term $\frac{1}{\alpha -1} \log 1/\eps$ in
\eqref{eqn:class-entropy-ineq} diverges when we try to recover the
Shannon entropy.  However, in the case of an i.i.d.\ sequence we find
\begin{equation}
\label{eqn:class-entropy-ineq-iid}
\frac{1}{n} \hmineps{\eps}{$X^n$}{}  \geq \halpha{$X$}{} - \frac{1}{n(1 - \alpha)} \log \eps,
\end{equation}
where we have used $\halpha{$X^n$}{} = n \halpha{$X$}{}$.
We proceed by bounding $\lim_{\eps \to 0} \lim_{n \to \infty} \hmineps{\eps}{$X^n$}{}$ from above and below.
To get the lower bound, we choose $\alpha = 1 + 1/\sqrt{n}$ and take the limit $n\! \to\! \infty$ in \eqref{eqn:class-entropy-ineq-iid}.
The upper bound essentially follows from $\hmin{$X$}{} \leq \hvn{$X$}{}$. 

\subsection{Fully Quantum AEP}

The AEP was first generalized
to situations where the outcomes A of the 
random experiment are quantum systems, while the side information
remains classical. In this case the side information
does not need to be modeled explicitly but can be included in the
description of the output states
(see e.g.\ \cite{schumacher94, barnum00, schoenmakers07, nielsen00}). 
In this paper we consider a generalization 
to a fully quantum AEP, 
involving possibly quantum mechanical side information B.\footnote{
We consider this result an extension of the AEP in the sense that it takes the role of the classical AEP in a quantum information context. Namely, if an information theoretic problem can be solved in terms of min- and max-entropies in a single-shot scenario, the asymptotic result for i.i.d.\ states follows via the AEP (and, thus, can be expressed in terms of von Neumann entropies). The classical AEP follows as a special case of our result.}
A preliminary version of the result 
has appeared in \cite{renner05} (see the discussion below for a comparison). Similar result have also been found in the context of quantum hypothesis testing~(e.g.~\cite{hiai91,ogawa00,hayashi06}).

\begin{theorem}[AEP]
\label{thm:qaep}
Let $\hA$ and $\hB$ be finite-dimensional Hilbert spaces, $\rhoAB$ a bipartite state on 
$\hA \kron \hB$ and $n \in \mathbb{N}$ s.t.~$\rhoAB^{\kron n}$ is an i.i.d.\ state on $(\hA
\kron \hB)^{\kron n}$, then
\begin{eqnarray}
&&\!\!\!\!\!\!\!\!\!\! \lim_{\eps \to 0} \lim_{n \to \infty} \frac{1}{n} \chmineps{\eps}{A$^n$}{B$^n$}{\rho^{\kron n}} = \chvn{A}{B}{\rho} \quad \textnormal{and} \quad 
\label{eqn:qaep1} \\
&&\!\!\!\!\!\!\!\!\!\! \lim_{\eps \to 0} \lim_{n \to \infty} \frac{1}{n} \chmaxeps{\eps}{A$^n$}{B$^n$}{\rho^{\kron n}} = \chvn{A}{B}{\rho}\, . \label{eqn:qaep2}
\end{eqnarray}
\end{theorem}
This relation is expressed in terms of quantum versions of the min-
and max-entropies \cite{renner05} and the conditional von Neumann
entropy which will be defined precisely below.
The reader unfamiliar with
quantum entropies is also referred to \cite{rennerkoenig05,koenig08} for
many of their properties and applications.  

In this contribution, we prove a non-asymptotic version of Theorem
\ref{thm:qaep} that gives a lower bound on
$\chmineps{\eps}{A$^n$}{B$^n$}{}$ for finite $n$ (cf.\ Theorem
\ref{thm:qep}).  The bound for finite $n$ has the property that the
deviation from the asymptotic bound 
(the term $\delta(\eps,\eta)$ in Theorem~\ref{thm:qep}) 
only depends on conditional min- and max-entropies
evaluated for $\rhoAB$ but is otherwise independent of the dimension
of the Hilbert spaces $\hA$ and $\hB$. More precisely, our bound is independent
of the Hilbert space dimension of $\hB$. This is particularly important
for applications in the context of cryptography, where quantum systems
may be controlled by an adversary. In this case, it is often difficult
or impossible to bound their dimension, whereas the conditional
entropies can nevertheless be estimated.

It is possible to obtain a statement similar to Theorem \ref{thm:qaep}
using typical subspaces. However, proofs of this type inevitably lead to
bounds involving the dimensions of both Hilbert spaces $\hA$ and $\hB$\footnote{
To our knowledge typical subspaces cannot be defined in a fully quantum setting. Instead, it is necessary
to first bound the conditional entropies in terms of (unconditional) entropies of the joint system AB and of system B separately. The typical subspace arguments can then be applied individually 
to get an asymptotic limit of $\hvn{AB}{} - \hvn{B}{} = \chvn{A}{B}{}$.
Bounds on the convergence derived from this argument will depend on the 
convergence of the individual terms and thus in general on the Hilbert space
dimensions of $\hAB$ and $\hB$ (see \cite{renner04}).}
and hence to a qualitatively weaker result than the one
established here (in particular, no reasonable
bound could be obtained for high-dimensional Hilbert spaces). We note
that our proof technique is different from the one used in
\cite{renner05}, where the result also explicitly depends on the dimension of the
Hilbert space $\hA$.

Our proof is based on quantities that
can be seen as a quantum generalization of R\'enyi entropies (Section \ref{sec:def}). A
central ingredient is a family of inequalities that generalize \eqref{eqn:class-entropy-ineq} to the
quantum domain (Section \ref{sec:ineq}). Together with a quantitative bound on the 
difference between the generalized R\'enyi entropies and the von Neumann entropy (Section \ref{sec:alpha-bound}), this leads to the main claim (Theorem \ref{thm:qep} in Section \ref{sec:qep}).

The technical tools used for the derivation of our results
(in particular Lemma \ref{lemma:monotonicity} in Appendix~\ref{app:tech}) may be of independent use\,|\,for
example, they allow for a simple proof of the strong sub-additivity of the
von Neumann entropy (cf. Lemma
\ref{lemma:dataproc} and \cite{hayashi06}).

\section{Quantum R\'enyi Entropies}
\label{sec:def}

In this section, we define the various entropies used and
explore some of their properties. Proofs of the lemmas can be found in Appendix \ref{app:prop}.
Given a finite-dimensional Hilbert space $\h$, we use $\positive{\h}$ to denote the set of positive semi-definite operators on $\h$.
The set of normalized quantum states $\states{\h} := \{ \rho \in \positive{\h} : \tr\,\rho = 1 \}$ and the set of sub-normalized states $\subnormstates{\h} := \{ \rho \in \positive{\h} : \tr\,\rho \leq 1 \}$ can now be defined.
Indices are used to denote multi-partite Hilbert spaces, e.g.\ $\hAB = \hA \kron \hB$.
Let  $\rhoAB \in \states{\hAB}$ be a bipartite state and $\sigmaB \in \states{\hB}$, then
\begin{equation*}
\chvn{A}{B}{\rho|\sigma} := \lim_{\xi \to 0} \Trace{\rhoAB (\idA \kron \log (\sigmaB \!+\! \xi \idB) - \log \rhoAB)}\, ,
\end{equation*}
where $\idA$ and $\idB$ are the identity operators on $\hA$ and $\hB$, respectively.
The conditional von Neumann entropy can then be recovered by
\begin{equation*}
\chvn{A}{B}{\rho} := \!\! \max_{\sigmaB \in \states{\hB}} \! \chvn{A}{B}{\rho|\sigma}  = \chvn{A}{B}{\rho|\rho}\, ,
\end{equation*}
where $\rhoB = \ptrace{A}{\rhoAB}$ is obtained by taking the partial trace on $A$ of $\rhoAB$.
We use indices to denote the different marginal states of multi-partite systems and often do not mention explicitly when a partial trace needs to be taken, since this information is contained implicitly in the notation of the entropies.\footnote{For example, given a state $\rhoABC$, the entropy $\chvn{A}{B}{\rho}$ is meant to be taken with the marginal states $\rhoAB = \ptrace{C}{\rhoABC}$ and $\rhoB = \ptrace{AC}{\rhoABC}$.} 
We define the min-entropy:
\begin{definition}
Let $\rhoAB \in \subnormstates{\hAB}$ and $\sigmaB \in \states{\hB}$, then
the min-entropy of A conditioned on B of the state $\rhoAB$ relative to $\sigmaB$ is defined as
\begin{equation}
\label{eqn:min-entropy-cond}
\chmin{A}{B}{\rho|\sigma} := \sup\, \{ \lambda \in \mathbb{R} : 2^{-\lambda}\, \idA \kron \sigmaB \geq \rhoAB \} 
\end{equation}
Furthermore, 
we define
\begin{equation}
\label{eqn:min-entropy}
\chmin{A}{B}{\rho} := \!\! \max_{\sigmaB \in \statesB} \! \chmin{A}{B}{\rho|\sigma} .
\end{equation}
\end{definition}
Clearly, $\chmin{A}{B}{\rho|\sigma}$ is finite if and only if $\supp{\sigmaB} \supseteq \supp{\rhoB}$
and $-\infty$ otherwise.
The max-entropy is its dual with regards to a purification $\rhoABC$
of $\rhoAB$ on an auxiliary Hilbert space
$\hC$:\footnote{\label{foot:max-entropy} Note that this quantity is
  different from the $H_\textrm{max}$ used in earlier work (e.g.\
  \cite{renner05}). However, the definition used here (and introduced
  in \cite{koenig08}) is chosen because it satisfies the duality
  relation \eqref{eqn:duality}. When $\hB \iso \mathbb{C}$, we recover the
  classical R\'enyi entropy $H_{\nicefrac{1}{2}}$. Note that the smooth 
  versions of $H_{\nicefrac{1}{2}}$ (cf.~eq.~\eqref{eqn:smoothmax}) and 
  $H_0$ are equivalent up to additive terms logarithmic in the smoothing
  parameter~\cite{renner04}, which disappear in the asymptotic statements.
   }
\begin{definition}
Let $\rhoABC \in \subnormstates{\hABC}$ be pure, then the max-entropy of A conditioned on B of
the state $\rhoAB$ is defined as
\begin{equation}
\label{eqn:duality}
\chmax{A}{B}{\rho} := -\chmin{A}{C}{\rho}\, .
\end{equation}
\end{definition} 
The quantum entropies can be ordered as follows:
\begin{lemma}
\label{lemma:ordering}
Let $\rhoAB \in \states{\hAB}$, then
\begin{equation}
\chmin{A}{B}{\rho} \leq \chvn{A}{B}{\rho} \leq \chmax{A}{B}{\rho} .
\label{eqn:ordering}
\end{equation}
\end{lemma}

In order to define smooth versions, we consider the set of states
close to $\rho$ in the following sense. For $\eps > 0$, we define an
$\eps$-ball of states around $\rho \in \states{\h}$ as
\begin{equation}
\epsball{\eps}{\rho} := \{ \rhot \in \subnormstates{\h} : C(\rho, \rhot) \leq \eps \} ,
\label{eqn:eps-ball}
\end{equation}
where $C(\rho, \rhot) := \sqrt{1 - F^2(\rho, \rhot)}$ as proposed in
\cite{nielsen04} is a distance measure (on normalized states) based on
the fidelity $F(\rho, \rhot) := \tr \abs{\sqrt{\rho}\sqrt{\rhot}}$. We
use this choice of measure because it is invariant under purifications
and is directly related to the trace distance for pure
states.\footnote{In fact, $C(\rho, \rhot)$ corresponds to the minimal
  trace distance between purifications of $\rho$ and $\rhot$ if
  $\trace{\rho} = \trace{\rhot} = 1$.}  Smoothed versions of the
min-entropy are then defined:
\begin{eqnarray*}
\chmineps{\eps}{A}{B}{\rho|\sigma} &:=& \max_{\rhotAB \in
  \epsball{\eps}{\rhoAB}} \chmin{A}{B}{\rhot|\sigma}\, ,\\
\chmineps{\eps}{A}{B}{\rho} &:=& \max_{\rhotAB \in
  \epsball{\eps}{\rhoAB}} \chmin{A}{B}{\rhot} \, .
\end{eqnarray*}
Similarly, we define
\begin{equation}
\label{eqn:smoothmax}
\chmaxeps{\eps}{A}{B}{\rho} := \min_{\rhotAB \in \epsball{\eps}{\rhoAB}} \chmax{A}{B}{\rhot}\, .
\end{equation}
The smoothed entropies maintain the duality relation
\begin{equation}
\label{eqn:eps-duality}
\chmaxeps{\eps}{A}{B}{\rho} = -\chmineps{\eps}{A}{C}{\rho} \, .
\end{equation}
Both entropies are independent of the Hilbert spaces used to represent
the density operators locally; namely, given $\rhoAB \in \states{\hAB}$, $\idx{\tau}{CD} \in \states{\hi{CD}}$ and
two isometries $U$ and $V$ s.t.\ $\idx{\tau}{CD} = (U\! \kron\! V) \rhoAB (U^\dagger\! \kron\! V^\dagger)$, we have
\begin{equation*} 
\chmineps{\eps}{A}{B}{\rho} = \chmineps{\eps}{C}{D}{\tau}\, , \ \ \chmaxeps{\eps}{A}{B}{\rho} = \chmaxeps{\eps}{C}{D}{\tau}\, .
\end{equation*}
Moreover, let $\sigmaB \in \states{\hB}$ and $\idx{\omega}{D}  := V \sigmaB V^\dagger$, then
\begin{equation}
\label{eqn:eps-iso}
\chmineps{\eps}{A}{B}{\rho|\sigma} = \chmineps{\eps}{C}{D}{\tau|\omega}\, .
\end{equation}
For a more in-depth treatment of smooth conditional entropies and their basic properties, we refer to \cite{tomamichel09}.

Next, we introduce a family of R\'enyi-like conditional entropies:
\begin{definition}
Let $\rhoAB \in \states{\hAB}$, $\sigmaB \in \statesB$ and $\alpha \in (0, 1) \cup (1, \infty)$, then the
$\alpha$-entropy of A conditioned on B of the state $\rhoAB$ relative to $\sigmaB$ is given by
\begin{equation}
\chalpha{A}{B}{\rho|\sigma} := \frac{ 1}{\!1 -\! \alpha} \log \trace{\rhoAB^\alpha\, \big(\idA \kron \sigmaB \big)^{1 - \alpha}}\, ,
\end{equation}
when $\sigmaB$ is invertible and $\lim_{\xi \to 0} \chalpha{A}{B}{\rho|\sigma+\xi \id}$ otherwise.
\end{definition}
Note that for $\alpha > 1$ the limit is finite if and only if $\supp{\sigmaB} \supseteq \supp{\rhoB}$
and $-\infty$ otherwise.
A similar quantity appears in quantum hypothesis testing
\cite{ogawa00,audenaert07} and as a quantum relative R\'enyi entropy
in \cite{hayashi06,ohya93,mosonyidatta08}.  If $\hB \iso \mathbb{C}$ is trivial,
we recover the classical R\'enyi entropies \eqref{eqn:renyi}. 
The entropies $H_0$ ($\alpha \!\to\! 0$) and $H_\infty$ ($\alpha \!\to\! \infty$) can be defined as limits. Moreover, the von Neumann entropy is recovered by continuous extension to $\alpha = 1$\,:
\begin{equation*}
\chx{1}{A}{B}{\rho|\sigma} := \lim_{\alpha \to 1} \chalpha{A}{B}{\rho|\sigma} = \chvn{A}{B}{\rho|\sigma}\, .
\label{eqn:renyi-to-vn}
\end{equation*}
Unlike their classical counterparts, the quantum conditional min-
and max-entropies cannot be recovered as special cases of
$\alpha$-entropies. However, it can be shown \cite{koenig08} that, for any
$\sigmaB \in \states{\hB}$,
\begin{equation}
\label{eqn:half-bound}
\chmax{A}{B}{\rho} = \!\!\! \max_{\idx{\tau}{B} \in
\states{\hB}}  \!\!\! \log F^2(\rhoAB, \idA \kron \idx{\tau}{B}) 
\geq \chx{\nicefrac{1}{2}}{A}{B}{\rho|\sigma}.
\end{equation}
Furthermore, using the eigenvalue decompositions $\rhoAB = \sum_{i}\nu_i\proj{i}{i}$ and
$\idA \kron \sigmaB =\sum_j\mu_j\proj{\bar{j}}{\bar{j}}$, we have
\begin{equation*}
\chx{\infty}{A}{B}{\rho|\sigma} = 
\lim_{\xi \to 0} -\log\max_{\genfrac{}{}{0pt}{}{i,j}{\braket{i}{\bar{j}}\neq 0}}\frac{\nu_i}{\mu_j+\xi}
\leq \chmin{A}{B}{\rho|\sigma}.
\end{equation*}

Nevertheless, the $\alpha$-entropies share some of the useful properties of their classical
counterparts:
\begin{lemma}
\label{lemma:mondec}
Let $\rhoAB \in \states{\hAB}$ and $\sigmaB \in \states{\hB}$, then
the entropies $\chalpha{A}{B}{\rho|\sigma}$ are monotonically decreasing in $\alpha$.
\end{lemma}

Furthermore, the entropies are additive, e.g.\ evaluation for an i.i.d.\ state $\rhoAB^{\kron n}$
relative to another i.i.d.\ state $\sigmaB^{\kron n}$ results in
\begin{equation}
\label{eqn:alpha-add}
\chalpha{A$^n$}{B$^n$}{\rho^{\kron n}|\sigma^{\kron n}} = n \chalpha{A}{B}{\rho|\sigma}\, .
\end{equation}
The $\alpha$-entropies are independent of the Hilbert spaces used
to represent the density operators locally:
\begin{lemma}
\label{lemma:alpha-iso}
Let $\rhoAB \in \states{\hi{AB}}$, $\idx{\sigma}{B} \in \states{\hi{B}}$, $\idx{\tau}{CD} \in \states{\hi{CD}}$, $\idx{\omega}{D} \in \states{\hi{D}}$ and
$U$, $V$ isometries s.t.\ $\idx{\tau}{CD} = {(U\!\kron\!V)} \rhoAB  (U^\dagger\!\kron\!V^\dagger)$ and $\idx{\omega}{D} =  V \sigmaB V^\dagger$. Then,
\begin{equation}
\label{eqn:alpha-iso}
\chalpha{A}{B}{\rho|\sigma} = \chalpha{C}{D}{\tau|\omega}\, .
\end{equation} 
\end{lemma}
Conditional entropies are measures of the uncertainty about $A$ given $B$, hence
we expect them to satisfy a data processing inequality, i.e.\ local processing
by a trace-preserving completely positive map (TP-CPM) on system B cannot 
decrease the conditional entropy. The $\alpha$-entropies for $\alpha \in [0, 2]$ have this property.
\begin{lemma}
\label{lemma:dataproc}
Let  $\alpha \in [0, 2]$, $\rhoAB \in \states{\hi{AB}}$, $\idx{\sigma}{B} \in \states{\hi{B}}$, $\idx{\tau}{AC} \in \states{\hi{AC}}$, $\idx{\omega}{C} \in \states{\hi{C}}$ and 
$\mathcal{E}$ a TP-CPM
s.t.\ $\idx{\tau}{AC} = \mathcal{I} \kron \mathcal{E} (\rhoAB)$ and $\idx{\omega}{C} = \mathcal{E} (\sigmaB)$, 
where $\mathcal{I}$ is the identity on $\states{\hA}$, then
\begin{equation}
\label{eqn:dataproc}
\chalpha{A}{B}{\rho|\sigma} \leq \chalpha{A}{C}{\tau|\omega}\, .
\end{equation}
\end{lemma} 
When the partial trace over a subsystem takes the role of the TP-CPM, Lemma~\ref{lemma:dataproc} is equivalent to strong sub-additivity in the case of the von Neumann entropy.
We find the following duality relation for $\alpha$-entropies:
\begin{lemma}
\label{lemma:alpha-dual} 
Let $\rhoABC \in \states{\hABC}$ be pure and $\alpha \in [0, 2]$, then
\begin{equation}
\label{eqn:alpha-dual}
\chalpha{A}{B}{\rho|\rho} = - \chx{2 - \alpha}{A}{C}{\rho|\rho} \, .
\end{equation}
\end{lemma}
The duality relation of the von Neumann entropy\,|\,
$\chvn{A}{B}{\rho} = -\chvn{A}{C}{\rho}$\,|\,follows
in the limit $\alpha \to 1$. 


\section{Lower Bound on Smooth Min-Entropy}
\label{sec:ineq}

Our main tool for proving the fully quantum AEP is a family of inequalities that relate the smooth conditional min-entropy 
$\chmineps{\eps}{A}{B}{}$ to $\chalpha{A}{B}{}$ for $\alpha \in (1, 2]$. 
The result is a quantum generalization of \eqref{eqn:class-entropy-ineq}.
\begin{theorem}
\label{thm:entropy-ineq}
Let $\rhoAB \in \states{\hAB}$, $\sigmaB \in \statesB$, $\eps > 0$ and $\alpha \in (1, 2]$, then the following inequality holds:
\begin{equation}
\label{eqn:entropy-ineq}
\chmineps{\eps}{A}{B}{\rho|\sigma} \geq \chalpha{A}{B}{\rho|\sigma} - \frac{1}{\alpha - 1} \log \frac{2}{\eps^2} .
\end{equation}
\end{theorem}

\begin{proof}
We consider two cases: (1) The $\alpha$ entropy diverges and the inequality holds trivially. (2) We have $\supp{\rhoB} \subseteq \supp{\sigmaB}$. In this case, we can find an isometry $\hB' \to \hB$ that maps a $\sigmaB'$ to $\sigmaB$ and $\rhoAB'$ to $\rhoAB$ s.t.\ $\sigmaB'$ has full support. The min- and $\alpha$-entropies are invariant under this isometry due to $\eqref{eqn:eps-iso}$ and Lemma~\ref{lemma:alpha-iso}, thus, we henceforth assume that $\sigmaB$ is invertible in this proof.

We use Appendix \ref{app:smooth-min} to get a first bound on $\chmineps{\eps}{A}{B}{}$; in particular, let $\lambda$ be chosen s.t.\ Lemma~\ref{lemma:smooth-min} holds for $\eps$ (cf.\ Remark~\ref{remark:choose-lambda}).
Next, we introduce the operator $X := \rhoAB - \lambda \idA \kron \sigmaB$ with eigenbasis $\{ \ket{\psi_i} \}_{i \in S}$. The set $S^+ \subseteq S$ contains the indices $i$ corresponding to positive eigenvalues of $X$. Hence, $P^+ := \sum_{i \in S^+}
\proj{\psi_i}{\psi_i}$ is the projector on the positive eigenspace
of $X$ and $P^+ X P^+=\Delta$ as defined in Lemma~\ref{lemma:smooth-min}. 
Furthermore, let $r_i :=
\bracket{\psi_i}{\rhoAB}{\psi_i} \geq 0$ and $s_i :=
\bracket{\psi_i}{\idA \kron \sigmaB}{\psi_i} > 0$. It follows that
\begin{equation*}
\forall_{i \in S^+}:\ r_i - \lambda s_i \geq 0 \quad \textnormal{and} \quad \frac{r_i}{\lambda\, s_i} \geq 1 .
\end{equation*}
For any $\alpha \in (1, 2]$, we bound $\eps$ in Lemma \ref{lemma:smooth-min} as follows:
\begin{eqnarray}
\frac{\eps^2}{2} &=& \trace{\Delta}
= \sum_{i \in S^+} r_i - \lambda s_i
\leq \sum_{i \in S^+} r_i \nonumber\\
&\leq& \sum_{i \in S^+} r_i \left( \frac{r_i}{\lambda\, s_i} \right)^{\alpha - 1} \nonumber\\
\label{eqn:eps-bound1}
&\leq& \lambda^{1 - \alpha}\, \sum_{i \in S} r_i^\alpha\, s_i^{1 - \alpha} . 
\end{eqnarray}

Next, we apply Lemma \ref{lemma:monotonicity} to the functional
$S_{g_\alpha}$, where $g_\alpha: t \mapsto t^\alpha$ is operator convex for $\alpha \in (1,
2]$ (cf.\ Section V.2 of \cite{bhatia97}).
We use the TP-CPM $A \mapsto \sum_{i \in S} \proj{\psi_i}{\psi_i} A \proj{\psi_i}{\psi_i}$ 
to obtain
\begin{equation*}
S_{g_\alpha}(\rhoAB, \idA \kron \sigmaB)=\trace{\rhoAB^\alpha (\idA \kron \sigmaB)^{1 - \alpha}} \geq \sum_{i \in S} r_i^\alpha s_i^{1 - \alpha} .
\end{equation*}
Substituting this into \eqref{eqn:eps-bound1}, we find
\begin{equation*}
\lambda^{\alpha - 1} \leq \frac{2}{\eps^2} \trace{\rhoAB^\alpha (\idA \kron \sigmaB)^{1 - \alpha}} \, .
\end{equation*}
Finally, taking the logarithm on both sides, dividing by $1 - \alpha < 0$ and applying Lemma \ref{lemma:smooth-min} results in \eqref{eqn:entropy-ineq}.
\end{proof}

\section{Lower Bound on $\alpha$-Entropies}
\label{sec:alpha-bound}

We will use Theorem \ref{thm:entropy-ineq} to get a lower bound on the min-entropy in terms of $\alpha$-entropies, hence, it remains to find a lower bound on the $\alpha$-entropies in terms of the von Neumann entropy. In turn, the bound on the convergence will depend on the smoothing parameter $\eps$ and a contribution $\Upsilon(\textrm{A}|\textrm{B})$ that describes how fast the $\alpha$-entropies converge to the von Neumann entropy.
\begin{definition}
Let $\rhoAB \in \states{\hAB}$ and $\sigmaB \in \states{\hB}$, then we define
the $\alpha$-entropy convergence parameter
\begin{equation}
\Upsilon(\textrm{A}|\textrm{B})_{\rho|\sigma} := 2^{-\frac{1}{2}\chx{\nicefrac{3}{2}}{A}{B}{\rho|\sigma}} + 2^{\frac{1}{2}\chx{\nicefrac{1}{2}}{A}{B}{\rho|\sigma}} + 1 \, .
\end{equation}
\end{definition}
When $\sigmaB = \rhoB$, one can use~\eqref{eqn:half-bound} and its dual relation $\chx{\nicefrac{3}{2}}{A}{B}{\rho|\rho} \geq \chmin{A}{B}{\rho}$ (cf.\ \eqref{eqn:duality} and Lemma~\ref{lemma:alpha-dual}) to write
$$ 
\Upsilon(\textrm{A}|\textrm{B})_{\rho|\rho} 
\leq \sqrt{2^{-\chmin{A}{B}{\rho}}} + \sqrt{2^{\chmax{A}{B}{\rho}}} + 1\, .$$
We can now state a bound 
on the $\alpha$-entropies for $\alpha$ close to $1$ 
as follows:
\begin{lemma}
\label{lemma:alpha-bound}
Let $\rhoAB \in \states{\hAB}$, $\sigmaB \in \states{\hB}$, $\eta = \Upsilon(\textrm{A}|\textrm{B})_{\rho|\sigma}$ and $1 < \alpha < 1 + \frac{\log 3}{4 \log \eta}$, then the following inequality holds:
\begin{equation}
\label{eqn:alphatohvn}
\chalpha{A}{B}{\rho|\sigma} 
\geq 
\chvn{A}{B}{\rho|\sigma} - 4\, (\alpha - 1) ( \log \eta )^2 \, .
\end{equation}
\end{lemma}

\begin{proof}
We assume that $\sigmaB$ is invertible in this proof. The general result then follows by the arguments outlined at the beginning of the proof of Theorem~\ref{thm:entropy-ineq}.


Let $\{ \ket{i} \}_i$ be an orthonormal basis of $\hAB$ and $\hAB' \iso \hAB$ a copy of $\hAB$. The state $\ket{\gamma} := \sum_i \ket{i} \kron \ket{i}$ is the (unnormalized) fully entangled state on $\hAB \kron \hAB'$. We introduce a purification $\ket{\phi} := (\sqrt{\rhoAB} \kron \idi{AB}) \ket{\gamma}$ of $\rhoAB$. To simplify notation, we use $\beta := \alpha - 1$ as well as $X := \rhoAB \kron (\id \kron \sigmaB^{-1})^T$. 

Let us first approximate $\chalpha{A}{B}{}$ for small $\beta > 0$:
\begin{equation*}
\chalpha{A}{B}{\rho|\sigma} = -\frac{1}{\beta} \log\, \bracket{\phi}{X^\beta}{\phi} \geq \frac{1}{\beta \ln 2} (1 - \bracket{\phi}{X^\beta}{\phi})\, ,
\end{equation*}
where we used $\ln x \leq x - 1$ for all $x > 0$.
We now expand the exponential $t^\beta$ for each eigenvalue $t > 0$ of $X$ as follows: $t^\beta = 1 + \beta \ln t + r_\beta(t)$,
where $r_\beta(t) := t^\beta - \beta \ln t - 1$. This leads to
\begin{eqnarray}
\label{eqn:lbound}
\chalpha{A}{B}{\rho|\sigma} &\geq& \frac{1}{\beta \ln 2} \big(- \beta \bracket{\phi}{\ln X}{\phi} - \bracket{\phi}{r_\beta(X)}{\phi} \big) \nonumber\\
&=& \chvn{A}{B}{\rho|\sigma} - \frac{1}{\beta \ln 2} \bracket{\phi}{r_\beta(X)}{\phi} .
\end{eqnarray}
To simplify this further, we note that
\begin{equation*}
r_\beta(t) \leq 2 (\cosh(\beta \ln t) - 1) =: s_\beta(t)\, .
\end{equation*}
It is easy to verify that $s_\beta$ is monotonically increasing for $t \geq 1$ and concave in $t$ for $\beta \leq 1/2$ and $t \in [3, \infty)$. 
Furthermore, we have $s_\beta(t) = s_\beta(\frac{1}{t})$ and $s_\beta(t^2) = s_{2\beta}(t)$. We use this to write\footnote{Adaptions of this step lead to different bounds. Here, we are interested in a bound that can be expressed in terms of $H_{\nicefrac{1}{2}}$ and $H_{\nicefrac{3}{2}}$.}
\begin{eqnarray}
s_\beta(t) &\leq& s_\beta \Big(t + \frac{1}{t} + 2 \Big)\ =\ s_{2\beta}\Big(\sqrt{t} + \frac{1}{\sqrt{t}} \Big) \nonumber\\
&\leq& s_{2\beta} \Big(\sqrt{t} + \frac{1}{\sqrt{t}} + 1 \Big) \label{eqn:s2beta} \, .
\end{eqnarray}
Next, we apply \eqref{eqn:s2beta} to the matrix element in \eqref{eqn:lbound} and use the fact that the operator $\sqrt{X} + 1/\sqrt{X} + \id$ has its eigenvalues in $[3, \infty)$ and $2 \beta < \frac{\log 3}{2 \log \eta} \leq \frac{1}{2}$ together with Lemma~\ref{lemma:jensen} in Appendix~\ref{app:tech}:
\begin{equation}
\label{eqn:rbsb}
\bracket{\phi}{r_\beta(X)}{\phi} \leq \bracket{\phi}{s_{2\beta}\Big(\sqrt{X} + \frac{1}{\sqrt{X}} + \id \Big)}{\phi} \leq s_{2\beta}(\eta)\, ,
\end{equation}
where we substituted $\eta = \bracket{\phi}{\sqrt{X} + 1/\sqrt{X} + \id}{\phi}$.
Taylor's theorem and an expansion around $\beta = 0$ gives an upper bound on $s_\beta(t)$:
$s_\beta(t) \leq \beta^2 (\ln t)^2 \cosh (\beta \ln t)$. Hence,
\begin{eqnarray}
\frac{1}{\beta \ln 2} s_{2\beta}(\eta)
&\leq&  \!\! 4 \beta (\log \eta)^2 \ln 2 \cosh(2 \beta \ln \eta) \nonumber\\
\label{eqn:s2bapprox}
&<&  \!\! 4 \beta (\log \eta)^2 \, ,
\end{eqnarray}
where we simplified the expression (for convenience of exposition) using $\ln 2 \cosh(\ln 3 / 2) < 1$. 
The lemma now follows after we substitute~\eqref{eqn:s2bapprox} and~\eqref{eqn:rbsb} into~\eqref{eqn:lbound}.
\end{proof}

\section{Quantum AEP}
\label{sec:qep}

One could use Theorem \ref{thm:entropy-ineq}, together with the
arguments given for the classical case in Section \ref{sec:classical},
to prove \eqref{eqn:qaep1} directly. In many applications, it is useful to
have an explicit lower bound on $\chmineps{\eps}{A}{B}{}$.
We derive such a bound, from which the
asymptotic version \eqref{eqn:qaep1} is a corollary.

\begin{theorem}
\label{thm:qep}
Let $\rhoAB \in \states{\hAB}$, $\eps > 0$, $\eta =  \Upsilon($A$|$B$)_{\rho|\rho}$ and $n \in \mathbb{N}$ s.t.\ $\rhoAB^{\kron n}$ is an i.i.d.\ state on $\hAB^{\kron n}$, then
\begin{equation*}
\frac{1}{n} \chmineps{\eps}{A$^n$}{B$^n$}{\rho^{\kron n}} 
\geq 
\chvn{A}{B}{\rho} - \frac{\delta(\eps, \eta)}{\sqrt{n}} \, ,
\end{equation*}
where, for $n \geq \frac{8}{5} \log \frac{2}{\eps^2}$, the error term is given by 
\begin{equation}
\label{eqn:errorterm}
\delta(\eps, \eta) := 4 \log \eta \sqrt{\log \frac{2}{\eps^2}} \, .
\end{equation}
\end{theorem}

\begin{proof}
By definition, we have
\begin{equation*}
\label{eqn:min-eps-rho}
\frac{1}{n} \chmineps{\eps}{A$^n$}{B$^n$}{\rho^{\kron n}} \geq
\frac{1}{n} \chmineps{\eps}{A$^n$}{B$^n$}{\rho^{\kron n}|\rho^{\kron n}} \, .
\end{equation*}
We use Theorem~\ref{thm:entropy-ineq}, the additivity property \eqref{eqn:alpha-add} of the $\alpha$-entropy and Lemma~\ref{lemma:alpha-bound} to get a bound on the rhs. Let $\alpha := 1 + \frac{1}{2 \mu \sqrt{n}}$ for a parameter $\mu$ (to be optimized over), then
\begin{eqnarray}
\textnormal{rhs.}\!\! &\geq&\!\! \frac{1}{n} \chalpha{A$^n$}{B$^n$}{\rho^{\kron n}|\rho^{\kron n}} - \frac{1}{n (\alpha - 1)} \log \frac{2}{\eps^2} \nonumber\\
&=&\!\! \chalpha{A}{B}{\rho|\rho} - \frac{2 \mu}{\sqrt{n}} \log \frac{2}{\eps^2} \nonumber\\
\label{eqn:mubound}
&\geq&\!\! \chvn{A}{B}{\rho|\rho} - \frac{2}{\sqrt{n}} \Big( \mu \log \frac{2}{\eps^2} + \frac{1}{\mu} (\log \eta)^2 \Big) .
\end{eqnarray}

We want to choose $\mu$ such that it minimizes the expression $\mu \log
\frac{2}{\eps^2} + \mu^{-1} (\log \eta)^2$. However, the requirement
$\alpha < 1 + \frac{\log 3}{4 \log \eta}$ in Lemma~\ref{lemma:alpha-bound} restricts the choice of $\mu$ for any fixed $n$, hence, the error term $\delta(\eps, \eta)$ is in general also a function of $n$. 
Nonetheless, for large enough $n$ the optimum, $\mu_*$, can be reached\footnote{To verify this, evaluate an upper bound to $\alpha = 1 + (2 \mu_* \sqrt{n})^{-1}$ using the expression for $n$ in~\eqref{eqn:mustar} and note that $\sqrt{5/2} < \log 3$.}  and we get
\begin{equation}
\label{eqn:mustar}
\mu_* = \sqrt{\frac{(\log \eta)^2}{\log \frac{2}{\eps^2}}} \quad \text{for} \quad n \geq \frac{8}{5} \frac{(\log \eta)^2}{\mu_*^{\,2}} = \frac{8}{5} \log \frac{2}{\eps^2} \, .
\end{equation}
Substitution of this expression into~\eqref{eqn:mubound} leads to~\eqref{eqn:errorterm}.
\end{proof}

\begin{remark}
\label{rmk:qep-extended}
The 
proof of Theorem~\ref{thm:qep} can be generalized to $\chmin{A}{B}{\rho|\sigma}$. The generalized theorem reads:
For $\rhoAB \in \states{\hi{AB}}$ and $\sigmaB \in \states{\hB}$, we have
$$\frac{1}{n} \chmineps{\eps}{A$^n$}{B$^n$}{\rho^{\kron n}|\sigma^{\kron n}} 
\geq
\chvn{A}{B}{\rho|\sigma} - \frac{\delta(\eps, \eta)}{\sqrt{n}} \, ,$$
where $n$ and $\delta(\eps, \eta)$ as in Theorem~\ref{thm:qep} and $\eta = \Upsilon($A$|$B$)_{\rho|\sigma}$.\footnote{Furthermore, note that all results of this paper can also be stated in terms of relative entropies instead of conditional entropies. For example, in the language of \cite{datta08v1}, Remark~\ref{rmk:qep-extended} reads: Let $\rho, \sigma \in \h$ and $\eps > 0$. With $n$ and $\delta$ as in Theorem~\ref{thm:qep}, we have $\frac{1}{n} D_\textrm{max}^\eps(\rho^{\kron n}\|\sigma^{\kron n}) \leq S(\rho\|\sigma) + \delta(\eps, \eta)/\sqrt{n}$, where the smoothing is over an $\eps$-ball around $\rho$ as defined in the present work and $\eta = 2^{\nicefrac{1}{2}\, S_{\nicefrac{3}{2}}(\rho\|\sigma)} + 2^{- \nicefrac{1}{2}\, S_{\nicefrac{1}{2}}(\rho\|\sigma)} + 1$.}
\end{remark}

The generalized asymptotic equipartition property stated in Theorem \ref{thm:qaep} follows as a corollary.
\begin{proof}[Proof of Theorem \ref{thm:qaep}]
We first show the min-entropy relation \eqref{eqn:qaep1}.
Taking the $n \to \infty$ limit in Theorem \ref{thm:qep} gives
\begin{equation*}
\lim_{\eps \to 0} \lim_{n \to \infty} \frac{1}{n} \chmineps{\eps}{A$^n$}{B$^n$}{\rho^{\kron n}} \geq \chvn{A}{B}{\rho}.
\end{equation*}

However, since, for $\eps \to 0$, the min-entropy is smaller than the
Shannon entropy (cf.\ Lemma \ref{lemma:ordering}), we get the desired
result. To see this, note that there exists a $\idx{\tilde{\rho}}{A$^n$B$^n$} \in \epsball{\eps}{\rhoAB^{\kron n}}$ s.t.\
\begin{equation}
\label{eqn:trivialbound}
\chmineps{\eps}{A$^n$}{B$^n$}{\rho^{\kron n}} = 
 \chmin{A$^n$}{B$^n$}{\tilde{\rho}} \leq \chvn{A$^n$}{B$^n$}{\tilde{\rho}} \, .
\end{equation}
We now use the continuity of the von Neumann entropy under small perturbations of the state as expressed in Fannes' inequality (cf.~\cite{alicki03}). This ensures that, for finite-dimensional
Hilbert spaces, the difference between the Shannon entropies evaluated for $\rhoAB^{\kron n}$ and $\idx{\tilde{\rho}}{A$^n$B$^n$}$ scales at most linearly in $n$ (i.e.\ logarithmic in the Hilbert space dimension):
\begin{equation*}
\lim_{n \to \infty} \frac{1}{n} \left| \chvn{A$^n$}{B$^n$}{\tilde{\rho}} - \chvn{A$^n$}{B$^n$}{\rho^{\kron n}} \right| \leq 
O(\eps) \, .
\end{equation*}
Together with~\eqref{eqn:trivialbound}, this leads to
\begin{equation*}
\lim_{\eps \to 0} \lim_{n \to \infty} \frac{1}{n} \chmineps{\eps}{A$^n$}{B$^n$}{\rho^{\kron n}} \leq \chvn{A}{B}{\rho}.
\end{equation*}

The max-entropy relation \eqref{eqn:qaep2} follows after we substitute
the duals of the smooth min-entropy \eqref{eqn:eps-duality} and the von Neumann entropy \eqref{eqn:alpha-dual} into \eqref{eqn:qaep1}.
\end{proof}

\appendix

\section{Technical Results}
\label{app:tech}

We discuss some properties of convex and operator convex\footnote{A continuous function $f$ on $[0, \infty)$ is operator convex if $f \big( \frac{1}{2}(A + B) \big) \leq \frac{1}{2} \big( f(A) + f(B) \big)$ for all positive semi-definite
matrices $A$ and $B$.} functions.
Let $\h$, $\h'$ be Hilbert spaces and $\ket{\phi} \in \h$. We start with a straightforward application of Jensen's inequality:

\begin{lemma}
\label{lemma:jensen}
Let $f$ be a convex function on $[a, b]$ and $X$ an operator on $\h$ s.t.\ $a\id \leq X \leq b\id$. Then, 
\begin{equation*}
\bracket{\phi}{f(X)}{\phi} \geq f(\bracket{\phi}{X}{\phi})\, .
\end{equation*}
\end{lemma}


A generalization of Jensen's inequality to operator convex functions was shown in \cite{hansen03}:
\begin{lemma}[Operator Jensen's Inequality]
\label{lemma:opjensen}
Let $f$ be a continuous operator convex function on $[0, \infty)$ and $\nu$ an isometry $\h \to \h'$. Then, for all $C \geq 0$ on $\h'$ it holds that 
$$\nu^\dagger f(C) \nu \geq f(\nu^\dagger C \nu) \, .$$
\end{lemma}

We will now discuss Lemma~\ref{lemma:monotonicity}, originally proven by Petz \cite{petz86}, which establishes
the monotonicity of certain functionals $S_f$ under TP-CPMs and is of independent use in quantum
information theory (see e.g.\ \cite{ogawa00, hayashi06}).
Let $\h$ be a Hilbert space with orthonormal basis $\{ \ket{i} \}_i$, $\h' \iso \h$ be a copy of $\h$, and $\ket{\gamma} := \sum_i \ket{i} \kron \ket{i}$ be the (unnormalized) fully entangled
state on $\h \kron \h'$. Then,
for any continuous function $f: [0, \infty) \to \mathbb{R}$ with $f(0) = 0$, 
and operators $A \geq 0, B > 0$ on $\h$, we define
\begin{equation*}
S_f(A, B) := \bracket{\gamma}{ (\sqrt{B} \kron \id) f(B^{-1} \kron A^{T}) (\sqrt{B} \kron \id)}{\gamma}\, ,
\end{equation*}
where $(\cdot)^T$ denotes the transpose with respect to $\{\ket{i} \}_i$ and $\id$ is the identity operator on $\h$.
More generally, for any $B \geq 0$, we define 
\begin{equation}
\label{eqn:Sfdefcont}
S_f(A, B) = \lim_{\xi \to 0} S_f(A, B + \xi \id) \, .
\end{equation}

The functional $S_f(A, B)$ is independent of the Hilbert space used to represent $A$ and $B$
in the following sense:
\begin{lemma}
\label{lemma:Sfiso}
Let $U: \h \to \bar{\h}$ be an isometry, then for all operators $A \geq 0$, $B \geq 0$ on $\h$:
$$ S_f(A, B) = S_f(U A U^\dagger, U B U^\dagger)\, .$$
\end{lemma}

\begin{proof}
Let $A = \sum_i \lambda_i \proj{\underline{i}}{\underline{i}}$ and $B = \sum_j \mu_j \proj{\bar{j}}{\bar{j}}$ with eigenvalues $\lambda_i \geq 0$, $\mu_j \geq 0$ and
orthonormal bases $\{ \ket{\underline{i}} \}_i$ and $\{ \ket{\bar{j}} \}_j$ respectively. Now, using $\bracket{\gamma}{X \kron \id}{\gamma} = \trace{X}$ 
and $(X \kron \id)\ket{\gamma} = (\id \kron X^T)\ket{\gamma}$ for any operator $X$, we get
\begin{equation*}
 S_f(A, B) = \lim_{\xi \to 0} \sum_{i,j} (\mu_j + \xi) f \left( \frac{\lambda_i}{\mu_j + \xi} \right) 
\abs{\braket{\underline{i}}{\bar{j}}}^2 \, .
\end{equation*}
The isometry $U$ keeps the eigenvalues and the scalar product $\braket{\underline{i}}{\bar{j}}$ invariant.
Furthermore, any zero eigenvalues introduced do not contribute to the sum since they lie on a space orthogonal to the image of $U$, and $f(0)=0$.
\end{proof}

\begin{lemma}
\label{lemma:monotonicity}
Let $f$ be operator convex
on $[0, \infty)$ and let
$\mathcal{E}$ be a TP-CPM, then for all operators $A \geq 0, B \geq 0$ on $\h$:
\begin{equation*}
S_f(A, B) \geq S_f(\mathcal{E}(A), \mathcal{E}(B))\, .
\end{equation*}
\end{lemma}

\begin{proof}
Every TP-CPM can be expressed as an isometry followed by a partial 
trace (cf.\ \cite{nielsen00}, Section 8.2). 
We have already established in Lemma \ref{lemma:Sfiso} that $S_f$ is invariant under isometries.
To show monotonicity under partial trace, we let $\h = \h_1 \kron \h_2$ with local bases
$\{ \ket{i}_1 \}_i$ and $\{ \ket{i}_2 \}_i$, respectively. We
introduce $\h_1' \iso \h_1$, the (unnormalized) fully entangled state
$\ket{\gamma} = \sum_{i,j} \ket{i}_1 \kron \ket{j}_2 \kron \ket{i}_1
\kron \ket{j}_2$ and its marginal 
$\ket{\gamma}_1 = \sum_i \ket{i}_1 \kron \ket{i}_1$.
It remains to show that $S_f(A, B) \geq S_f(A_1, B_1)$, where $A_1 = \ptrace{2}{A}$ and $B_1 = \ptrace{2}{B}$. We will show
this under the assumption that $B$ is invertible and the result for general $B$ will follow
from the continuity (by definition) of $S_f$ when $\xi \to 0$.

Let us define a linear map $\nu: \h_1 \kron \h_1' \to \h \kron \h'$ by
\begin{equation*}
\nu := \sum_i 
\left( 
\sqrt{B} \left( \sqrt{B_1}^{\,-1} \! \kron \ket{i}_2 \right) 
\right) 
\kron \id_1 \kron \ket{i}_2 \, .
\end{equation*}
The map $\nu$ is an isometry, i.e.\ $\nu^\dagger \nu = \id_{11}$ and satisfies 
\begin{equation}
\label{eqn:iso-on-sqrt}
\nu \big( \sqrt{B_1} \kron \id_1 \ket{\gamma}_1 \big)  = \sqrt{B} \kron \id_{12} \ket{\gamma}\, .
\end{equation}
Moreover, we have $\ptrace{2}{A^T} = A_1^T$, since the transpose is taken in the product basis. Hence, it follows that 
\begin{equation*}
\nu^\dagger (B^{-1} \kron A^T) \nu = B_1^{-1} \kron A_1^T \, .
\end{equation*}
Next, we apply Lemma~\ref{lemma:opjensen} to get
\begin{equation*}
\nu^\dagger f(B^{-1} \kron A^T) \nu \geq f \big( \nu^\dagger (B^{-1} \kron A^T) \nu )
= f(B_1^{-1} \kron A_1^T)\, . 
\end{equation*}
Finally, using \eqref{eqn:iso-on-sqrt}, we recover $S_f(A, B) \geq S_f(A_1, B_1)$
by taking the matrix element for $(\sqrt{B_1} \kron \id_1 ) \ket{\gamma}_1$ on both sides of the inequality.
\end{proof}

\section{Proofs of claims in Section \ref{sec:def}}
\label{app:prop}

For completeness, we prove various properties of the min-, max- and $\alpha$-entropies presented in Section \ref{sec:def}.
  
\begin{proof}[Proof of Lemma \ref{lemma:ordering}]
It is sufficient to prove the first relation\footnote{See also Lemma 10 in \cite{datta08v1} for an alternative
    proof.  There they define the relative entropy $D_{\textnormal{max}}(\rhoB \| \sigmaB)$
    which, for $\hA \iso \mathbb{C}$ trivial, is equal to $-\chmin{A}{B}{\rho|\sigma}$.}, since $\chmin{A}{B}{} \leq \chvn{A}{B}{}$ implies
$\chmax{A}{C}{} \geq \chvn{A}{C}{}$ by the duality relations.
\begin{eqnarray*}
\chvn{A}{B}{\rho} &=& \max_{\sigmaB \in \statesB} \tracebig{ \rhoAB (\log (\idA \kron \sigmaB) - \log \rhoAB ) } \\
&\geq& \Trace{\rhoAB ( \log(\lambda \idA \kron \sigmaB') - \log \rhoAB )} - \log \lambda \\
&\geq& \chmin{A}{B}{\rho}\ ,
\end{eqnarray*}
where we chose $\lambda > 0$ and $\sigmaB' \in \statesB$ such that they optimize~\eqref{eqn:min-entropy-cond} and~\eqref{eqn:min-entropy}. Hence, $-\log \lambda = \chmin{A}{B}{\rho}$. Furthermore, it follows from~\eqref{eqn:min-entropy-cond} that $\lambda \idA \kron \sigmaB' \geq \rhoAB$. Then, using the operator monotonicity\footnote{A function $f$ on $(0, \infty)$ is operator monotone if $A \geq B$ implies $f(A) \geq f(B)$ for any strictly positive Hermitian matrices $A$ and $B$.} of $t \mapsto \log t$ (cf.\ Chapter V in \cite{bhatia97}), we find that the remaining term is positive.
\end{proof}

\begin{proof}[Proof of Lemma \ref{lemma:mondec}]
We prove this statement for invertible $\sigmaB$ and the general statement then follows by continuity. Using the (unnormalized) fully entangled state $\ket{\gamma}$ as in Appendix~\ref{app:tech}, we define a purification $\ket{\phi} := (\sqrt{\rhoAB} \kron \idi{AB}) \ket{\gamma}$ of $\rhoAB$. Furthermore, we set $\beta := \alpha - 1$ and $X := \rhoAB \kron (\id \kron \sigmaB^{-1})^T$. It is easy to verify that, for $f: t \mapsto t \log t$,
\begin{eqnarray*}
\chalpha{A}{B}{\rho|\sigma} &=& -\frac{1}{\beta} \log\, \bracket{\phi}{X^\beta}{\phi}\, \quad \textnormal{and} \\
\frac{\partial}{\partial \alpha} \chalpha{A}{B}{\rho|\sigma} &=& \frac{1}{\beta^2} \log\, \bracket{\phi}{X^\beta}{\phi} - \frac{1}{\beta} \frac{\bracket{\phi}{X^\beta \log X}{\phi}}{\bracket{\phi}{X^\beta}{\phi}} \\
&=& \frac{ f(\bracket{\phi}{X^\beta}{\phi}) - \bracket{\phi}{f(X^\beta)}{\phi} }{\beta^2\, \bracket{\phi}{X^\beta}{\phi} } \, .
\end{eqnarray*}
The statement of the lemma then follows from the convexity of $f$ together with Lemma~\ref{lemma:jensen}.
\end{proof}

\begin{proof}[Proof of Lemma \ref{lemma:alpha-iso}]
The $\alpha$-entropies can be expressed in terms of the functionals of Appendix~\ref{app:tech}. 
Given the continuous functions $g_\alpha: t \mapsto t^\alpha$ and $h: t \mapsto - t \log t$ that satisfy $g_\alpha(0) = 0$ for $\alpha \in (0, \infty)$ and $h(0) = 0$, we write
\begin{equation}
\label{eqn:alpha-alt-def}
\chalpha{A}{B}{\rho|\sigma} = \left\{ \begin{array}{ll} 
\!\!\frac{1}{1\!-\!\alpha}\! \log S_{g_\alpha}(\rhoAB, \idA \kron \sigmaB) & \alpha \in (0, 1) \\
S_{h}(\rhoAB, \idA \kron \sigmaB) & \alpha = 1 \\
\!\!-\frac{1}{\alpha\!-\!1}\! \log S_{g_\alpha}(\rhoAB, \idA \kron \sigmaB) & \alpha \in (1, \infty) \\
\end{array} \right.  .
\end{equation}
The proof is now a straightforward application of Lemma~\ref{lemma:Sfiso} with isometry $U \kron V$ to
the functionals $S_h$ and $S_{g_\alpha}$. Note that
$(U\!\kron\!V)(\idA\!\kron\!\sigmaB) (U^\dagger\!\kron\!V^\dagger) = U U^\dagger \kron \idx{\omega}{D}$, where
$U U^\dagger$ and $\idi{C}$ are interchangeable in the definition of $\chalpha{C}{D}{\tau|\omega}$, since $\idx{\tau}{C}$ has its support on $U U^\dagger$. The statements for $\alpha \to 0$ and $\alpha \to \infty$ follow by continuity.
\end{proof}

\begin{proof}[Proof of Lemma \ref{lemma:dataproc}]
The proof is a straightforward application of Lemma~\ref{lemma:monotonicity} with TP-CPM $\mathcal{I} \kron \mathcal{E}$
to the functionals in~\eqref{eqn:alpha-alt-def}.
In the von Neumann limit $\alpha = 1$, we write 
\begin{eqnarray*}
\chvn{A}{B}{\rho|\sigma} &=& S_h(\rhoAB, \idA \kron \idx{\sigma}{B}) \\
&\leq& S_h( \mathcal{I} \kron \mathcal{E}(\rhoAB),  \mathcal{I} \kron \mathcal{E} (\idA \kron \sigmaB) ) \\
&=& S_h(\idx{\tau}{AC}, \idA \kron \idx{\omega}{C})\ =\ \chvn{A}{C}{\tau|\omega} \, ,
\end{eqnarray*}
where we used that $h: t \mapsto - t \log t$ is operator concave on $[0, \infty)$ (cf.\ Chapter V in \cite{bhatia97}). 
Similarly, we use that $g_\alpha: t \mapsto t^\alpha$ on $[0, \infty)$ is operator
concave for $\alpha \in (0, 1)$ and operator convex for $\alpha \in (1, 2]$ (cf.\ Chapter V in \cite{bhatia97}) as well
as continuity at $\alpha = 0$ to prove the statement for $\alpha \in [0, 2]$.
\end{proof}

\begin{proof}[Proof of Lemma \ref{lemma:alpha-dual}]
We write $\rhoABC = \proj{\vartheta}{\vartheta}$ and note that the marginal states $\rhoAB$ and $\idx{\rho}{C}$ satisfy
$(\rhoAB \kron \idi{C}) \ket{\vartheta} = (\idi{AB} \kron \idx{\rho}{C}) \ket{\vartheta}$. The same applies to $\rhoB$ and $\idx{\rho}{AC}$. Thus,
\begin{eqnarray*}
&& \!\!\!\! (1\! -\! \alpha) \chx{\alpha}{A}{B}{\rho|\rho} \\
&&\quad =\ \log \trace{\rhoAB^\alpha (\idA \kron \rhoB)^{1 - \alpha}} \\
&&\quad =\ \log \bracket{\vartheta}{(\rhoAB \kron \idi{C})^{\alpha - 1} (\idA \kron \rhoB \kron \idi{C})^{1 - \alpha}}{\vartheta} \\
&& \quad =\ \log \trace{ (\idA \kron \rhoC)^{\alpha - 1} \rhoAC^{2 - \alpha} } \, .
\end{eqnarray*}
The equality now follows from $\alpha - 1 =  1 - (2 - \alpha)$.
\end{proof}

\section{Estimate of $\chmineps{\eps}{A}{B}{\rho}$}
\label{app:smooth-min}

The following lemma gives an estimate of the smooth min-entropy
$\chmineps{\eps}{A}{B}{\rho}$ (see also~\cite{dattarenner08}):
\begin{lemma}
\label{lemma:smooth-min}
Let $\rhoAB \in \states{\hAB}$, $\sigmaB \in \statesB$ and $\lambda > 0$, then
\begin{equation}
\chmineps{\eps}{A}{B}{\rho|\sigma} \geq - \log \lambda, \quad \eps = \sqrt{2\, \trace{\Delta}} ,
\label{eqn:smooth-min-approx}
\end{equation}
where $\Delta := \{ \rhoAB - \lambda \idA \kron \sigmaB \}_+$ is the positive part of the operator $\rhoAB - \lambda \idA \kron \sigmaB$.
\end{lemma}

\begin{proof}
We first choose $\rhotAB$, bound $\chmin{A}{B}{\tilde{\rho}|\sigma}$, and then show that 
$\rhotAB \in \epsball{\eps}{\rhoAB}$. We use the 
abbreviated notation $\Lambda := \lambda \idA \kron \sigmaB$ and set
\begin{equation*}
\rhotAB := G \rhoAB G^\dagger, \qquad G := {\Lambda}^{\frac{1}{2}}\, (\Lambda + \Delta)^{-\frac{1}{2}}\, ,
\end{equation*}
where the inverse is taken on the support of $\Lambda$. 
From the definition of $\Delta$, we have $\rhoAB \leq \Lambda + \Delta$;
hence, $\rhotAB \leq \Lambda$ and $\chmin{A}{B}{\tilde{\rho}|\sigma} \geq - \log \lambda$.

Let $\ket{\psi}$ be a purification of $\rhoAB$, then $(G \kron \idi{AB}) \ket{\psi}$ is a purification of $\rhotAB$ and, using Uhlmann's theorem, we find a bound on the fidelity:
\begin{equation*}
F(\rhoAB, \rhotAB) \geq \abs{\bracket{\psi}{G \kron \idi{AB}}{\psi}} \geq \Re \big\{ \trace{G\rhoAB} \big\} = \trace{\bar{G}\rhoAB}\, ,
\end{equation*}
where we introduced $\bar{G} := \frac{1}{2}(G + G^\dagger)$.
Hence,
\begin{equation*}
C(\rhoAB, \rhotAB) \leq \sqrt{(1 + \trace{\bar{G}\rhoAB})(1 - \trace{\bar{G}\rhoAB})} \, .
\end{equation*}
This can be simplified further after we note that $G$ is a contraction.\footnote{A contraction $G$ is an operator with operator norm $\norm{G}{} \leq 1$.} 
To see this, we multiply $\Lambda \leq \Lambda + \Delta$ with $(\Lambda + \Delta)^{-\frac{1}{2}}$ from
left and right to get
\begin{equation*}
G^\dagger G = (\Lambda + \Delta)^{-\frac{1}{2}} \Lambda (\Lambda + \Delta)^{-\frac{1}{2}} \leq \idi{AB}.
\end{equation*}
Furthermore, $\bar{G} \leq \idi{AB}$, since $\norm{\bar{G}}{} \leq 1$ by the triangle inequality and $\norm{G}{} = \norm{G^\dagger}{} \leq 1$. Clearly, $\trace{\bar{G}\rhoAB} \leq 1$. Moreover,
\begin{eqnarray*}
1 - \trace{\bar{G}\rhoAB} &=& \trace{(\idi{AB} - \bar{G})\rhoAB} \\
&\leq& \trace{\Lambda + \Delta} - \trace{\bar{G} (\Lambda + \Delta)} \\
&=& \trace{\Lambda + \Delta} - \trace{(\Lambda + \Delta)^{\nicefrac{1}{2}} {\Lambda}^{\nicefrac{1}{2}} }\\
&\leq& \trace{\Delta}\, ,
\end{eqnarray*}
where we used $\rhoAB \leq \Lambda + \Delta$ and $\sqrt{\Lambda + \Delta} \geq \sqrt{\Lambda}$. The latter inequality follows from the operator monotonicity of the square root function (Proposition V.1.8 in \cite{bhatia97}).
Finally, $C(\rhoAB, \rhotAB) \leq \sqrt{2\trace{\Delta}} = \eps$ and $\rhotAB \in \epsball{\eps}{\rhoAB}$. 
\end{proof}

\begin{remark}
\label{remark:choose-lambda}
For a fixed $\eps \in [0, 1]$ and $\supp{\rhoB} \subseteq \supp{\sigmaB}$, we can always find a finite 
$\lambda$ s.t.~Lemma~\ref{lemma:smooth-min} holds. To see this, note that
$$\eps(\lambda) = \sqrt{ 2\, \tr\, \{ \rhoAB - \lambda \idA \kron \sigmaB \}_+ }$$ 
is continuous in $\lambda$ with $\eps(0) = \sqrt{2}$ and $\eps(2^{-\chmin{A}{B}{\rho|\sigma}}) = 0$.
\end{remark}

\section*{Acknowledgments}

We thank Johan \AA berg, Nilanjana Datta, Mil\'an Mosonyi and J\"urg Wullschleger for comments.
We acknowledge support from the Swiss National Science Foundation (grant No. 200021-119868).


\begin{thebibliography}{10}
\providecommand{\url}[1]{#1}
\csname url@samestyle\endcsname
\providecommand{\newblock}{\relax}
\providecommand{\bibinfo}[2]{#2}
\providecommand{\BIBentrySTDinterwordspacing}{\spaceskip=0pt\relax}
\providecommand{\BIBentryALTinterwordstretchfactor}{4}
\providecommand{\BIBentryALTinterwordspacing}{\spaceskip=\fontdimen2\font plus
\BIBentryALTinterwordstretchfactor\fontdimen3\font minus
  \fontdimen4\font\relax}
\providecommand{\BIBforeignlanguage}[2]{{%
\expandafter\ifx\csname l@#1\endcsname\relax
\typeout{** WARNING: IEEEtran.bst: No hyphenation pattern has been}%
\typeout{** loaded for the language `#1'. Using the pattern for}%
\typeout{** the default language instead.}%
\else
\language=\csname l@#1\endcsname
\fi
#2}}
\providecommand{\BIBdecl}{\relax}
\BIBdecl

\bibitem{cover91}
T.~M. Cover and J.~A. Thomas, \emph{Elements of Information Theory}.\hskip 1em
  plus 0.5em minus 0.4em\relax Wiley Series in Telecommunications, 1991.

\bibitem{holenstein06}
\BIBentryALTinterwordspacing
T.~Holenstein and R.~Renner, ``On the randomness of independent experiments,''
  2006. [Online]. Available: \url{http://arxiv.org/abs/cs/0608007}
\BIBentrySTDinterwordspacing

\bibitem{renner04}
R.~Renner and S.~Wolf, ``Smooth {R}{\'e}nyi entropy and applications,'' in
  \emph{Proc. ISIT, 2004}.

\bibitem{renyi61}
A.~R\'enyi, ``On measures of information and entropy,'' in \emph{Proc. 4th
  Berkeley Symp. on Math., Stat. and Prob., 1961}, pp. 547--561.

\bibitem{schumacher94}
R.~Jozsa and B.~Schumacher, ``A new proof of the quantum noiseless coding
  theorem,'' \emph{J. Mod. Opt.}, vol.~41, pp. 2343--2349, 1994.

\bibitem{barnum00}
H.~Barnum, E.~Knill, and M.~Nielsen, ``On quantum fidelities and channel
  capacities,'' \emph{IEEE Trans. Inf. Theory}, vol.~46, no.~4, pp. 1317--1329,
  Jul. 2000.

\bibitem{schoenmakers07}
B.~Schoenmakers, J.~Tjoelker, P.~Tuyls, and E.~Verbitskiy, ``Smooth {R}{\'e}nyi
  entropy of ergodic quantum information sources,'' in \emph{Proc. ISIT, 2007},
  pp. 256--260.

\bibitem{nielsen00}
M.~A. Nielsen and I.~L. Chuang, \emph{Quantum Computation and Quantum
  Information}.\hskip 1em plus 0.5em minus 0.4em\relax Cambridge University
  Press, 2000.

\bibitem{renner05}
\BIBentryALTinterwordspacing
R.~Renner, ``Security of quantum key distribution,'' Ph.D. dissertation, Swiss
  Federal Institute of Technology, Zurich, 2005. [Online]. Available:
  \url{http://arxiv.org/abs/quant-ph/0512258}
\BIBentrySTDinterwordspacing

\bibitem{hiai91}
F.~Hiai and D.~Petz, ``The proper formula for relative entropy and its
  asymptotics in quantum probability,'' \emph{Commun. Math. Phys.}, vol. 143,
  pp. 99--114, 1991.

\bibitem{ogawa00}
T.~Ogawa and H.~Nagaoka, ``Strong converse and {S}tein's lemma in quantum
  hypothesis testing,'' \emph{IEEE Trans. Inf. Theory}, vol.~46, pp.
  2428--2433, Nov. 2000.

\bibitem{hayashi06}
M.~Hayashi, \emph{Quantum Information --- An Introduction}.\hskip 1em plus
  0.5em minus 0.4em\relax Springer, 2006.

\bibitem{rennerkoenig05}
\BIBentryALTinterwordspacing
R.~Renner and R.~K\"onig, ``Universally composable privacy amplification
  against quantum adversaries,'' in \emph{Proc. TCC, 2005}, pp. 407--425.
  [Online]. Available: \url{http://arxiv.org/abs/quant-ph/0403133}
\BIBentrySTDinterwordspacing

\bibitem{koenig08}
\BIBentryALTinterwordspacing
R.~K\"onig, R.~Renner, and C.~Schaffner, ``The operational meaning of min- and
  max-entropy,'' Jul. 2008. [Online]. Available:
  \url{http://arxiv.org/abs/0807.1338}
\BIBentrySTDinterwordspacing

\bibitem{nielsen04}
\BIBentryALTinterwordspacing
A.~Gilchrist, N.~K. Langford, and M.~A. Nielsen, ``Distance measures to compare
  real and ideal quantum processes,'' Aug. 2004. [Online]. Available:
  \url{http://arxiv.org/abs/quant-ph/0408063}
\BIBentrySTDinterwordspacing

\bibitem{tomamichel09}
M.~Tomamichel, R.~Colbeck, and R.~Renner, ``On the smoothing of conditional
  min- and max-entropies,'' Unpublished Notes, 2009.

\bibitem{audenaert07}
K.~M.~R. Audenaert, J.~Calsamiglia, R.~Munoz-Tapia, E.~Bagan, L.~Masanes,
  A.~Acin, and F.~Verstraete, ``Discriminating states: The quantum chernoff
  bound,'' \emph{Phys. Rev. Lett.}, vol.~98, no.~16, p. 160501, 2007.

\bibitem{ohya93}
M.~Ohya and D.~Petz, \emph{Quantum Entropy and Its Use}.\hskip 1em plus 0.5em
  minus 0.4em\relax Springer, 1993.

\bibitem{mosonyidatta08}
\BIBentryALTinterwordspacing
M.~Mosony and N.~Datta, ``Generalized relative entropies and the capacity of
  classical-quantum channels,'' 2008. [Online]. Available:
  \url{http://arxiv.org/abs/0810.3478}
\BIBentrySTDinterwordspacing

\bibitem{bhatia97}
R.~Bhatia, \emph{Matrix Analysis}, ser. Graduate Texts in Mathematics.\hskip
  1em plus 0.5em minus 0.4em\relax Springer, 1997.

\bibitem{datta08v1}
\BIBentryALTinterwordspacing
N.~Datta, ``Min- and max- relative entropies and a new entanglement monotone,''
  Mar. 2008. [Online]. Available: \url{http://arxiv.org/abs/0803.2770v1}
\BIBentrySTDinterwordspacing

\bibitem{alicki03}
\BIBentryALTinterwordspacing
R.~Alicki and M.~Fannes, ``Continuity of quantum mutual information,'' Dec.
  2003. [Online]. Available: \url{http://arxiv.org/abs/quant-ph/0312081v1}
\BIBentrySTDinterwordspacing

\bibitem{hansen03}
F.~Hansen and G.~K. Pedersen, ``Jensen's operator inequality,'' \emph{Bull.
  London Math. Soc.}, vol.~35, no.~5, pp. 553--564, Jul. 2003.

\bibitem{petz86}
D.~Petz, ``Quasi-entropies for finite quantum systems,'' \emph{Rep. Math.
  Phys.}, vol.~23, no.~1, pp. 57--65, Sep. 1984.

\bibitem{dattarenner08}
\BIBentryALTinterwordspacing
N.~Datta and R.~Renner, ``Smooth {R}{\'e}nyi entropies and the quantum
  information spectrum,'' Jan. 2008. [Online]. Available:
  \url{http://arxiv.org/abs/0801.0282}
\BIBentrySTDinterwordspacing

\end{thebibliography}
\end{document}